\documentclass[journal=jacsat,manuscript=article]{achemso}

\usepackage{chemformula} 
\usepackage[T1]{fontenc} 
\usepackage{braket}
\usepackage{commath}
\usepackage{graphicx}

\author{Giacomo Mariani}
 \affiliation{%
 Division of Physics, University of Tsukuba, Tennodai, Tsukuba, Ibaraki 305-8571, Japan
 }
  \author{Shuhei Nomoto}%
  \affiliation{%
 Division of Physics, University of Tsukuba, Tennodai, Tsukuba, Ibaraki 305-8571, Japan
 }

\author{Satoshi Kashiwaya}%
 \affiliation{%
Department of Applied Physics, Chikusa-Ku, Nagoya 464-8571, Japan
 }
 \author{Shintaro Nomura}%
 \email{nomura.shintaro.ge@u.tsukuba.ac.jp}
\affiliation{%
 Division of Physics, University of Tsukuba, Tennodai, Tsukuba, Ibaraki 305-8571, Japan
 }
\title{Imaging of microwave field distribution\\  over a non-fed gold pattern\\ by using NV centers in diamond}
\begin{document}
\clearpage
\begin{abstract}
Nitrogen-vacancy (NV) centers in diamond have been widely used as platforms for quantum information, magnetometry and imaging of microwave (MW) fields. High-precision spatial control of the MW field necessary to drive the electronic spin of NV centers is essential for these applications. Here, we report a controlled MW field distribution by excitation of a micrometer-scale gold pattern in vicinity of the diamond surface. The gold pattern excited by a planar ring MW antenna, acts as a receiving antenna and redistribute the MW field in a localized area, without a direct feed of electrical current. The planar ring MW antenna is designed to generate a uniform MW field on diamond substrate in an area of 0.785 mm$^{2}$, providing a useful tool for detecting the MW variations. We performed the imaging of the localized MW intensity on the micrometer-scale gold pattern by direct observation of electron spin Rabi oscillations, showing also the potential application of NV centers for imaging MW field and characterization of MW devices. We achieved an enhancement of about 19 times for the Rabi frequency on a scale of few micrometers for the gold pattern, compared to the bulk Rabi frequency in presence of the single planar ring MW antenna. Compared to previous methods, our method has been shown as a fast and easy tool for the spatial control of MW fields and spin manipulation of NV centers in diamond.
\end{abstract}
\clearpage
\section{}
The coherent manipulation of the electron and nuclear spin in nitrogen-vacancy (NV) centers in diamond has become fundamental for both quantum information processing and sensing applications \cite{DOHERTY20131, Degen2017}. Single and double quantum bits (qubits) are realized in NV centers by driving single spins with resonant microwave (MW) or radio-frequency (RF) fields and specific pulse sequences \cite{Jelezko2004b, Childress281, Dutt1312, Maurer1283}. Ensemble of NV centers have proved to be excellent magnetometers \cite{Taylor2008, Maertz2010, Simpson2016, Chang2017, Miura2017} offering high-spatial resolution with a signal-noise ratio proportional to $\sqrt{N_{NV}}$, where $N_{NV}$ is the number of the driven spins. For these applications, a precise spatial control of the MW field distribution is fundamental to drive coherently single or ensembles of spin. In imaging and sensing applications, the commonly employed MW antennas have a large bandwidth and can generate uniform MW fields in a wide area \cite{Bayat2014, Mrozek2015, Zhang2018} or in a 3D volume \cite{Chen2018, Kapitanova2018}. For quantum information, the area of interest is limited to few or single centers, for which high and more localized MW magnetic fields are preferred to the drive spin in efficient way. For instance, miniaturized MW loops \cite{Jelezko2004} thin wires and coplanar waveguides \cite{Fuchs1520} fabricated directly on the diamond surface can offer higher magnetic field amplitude in the near-field but they can be easily subjected to disconnections or induce undesired sample heating.

Here, we demonstrated a simple method to redistribute the MW magnetic field on a diamond substrate by resonant excitation of a micrometer-scale gold loop-pattern coated on silicon. By using an ensemble of near-surface NV centers in the same diamond substrate, we performed the imaging of the MW magnetic field distribution in the near-field. In our experimental setup, we use a large MW antenna which provides a uniform MW field over the sample at a distance of about 0.5 mm. The micrometer-scale gold pattern in close proximity with the diamond surface, acts as a resonant antenna and redistributes the MW field, providing an enhanced field in a localized area; the remitted MW field is then measured by NV centers. Ensemble of NV centers have been used for imaging microwave magnetic fields previously \cite{Appel2015, Horsley2018,Yang2018, Wang2018}; by driving the electron spin with the MW field, we measured the frequency of Rabi oscillations which is directly associated to the microwave intensity. This method gives a quantitatively accurate measure of the MW intensity. Compared to previous excitation methods, the micrometer-scale gold pattern is not fed by electrical current and it's shape can be design to create a specific MW magnetic field distribution on the diamond. This structure is an easy and fast way to create ad-hoc MW distributions for specific applications and limiting MW power losses.

\begin{figure}[!ht]
\includegraphics[width=\linewidth]{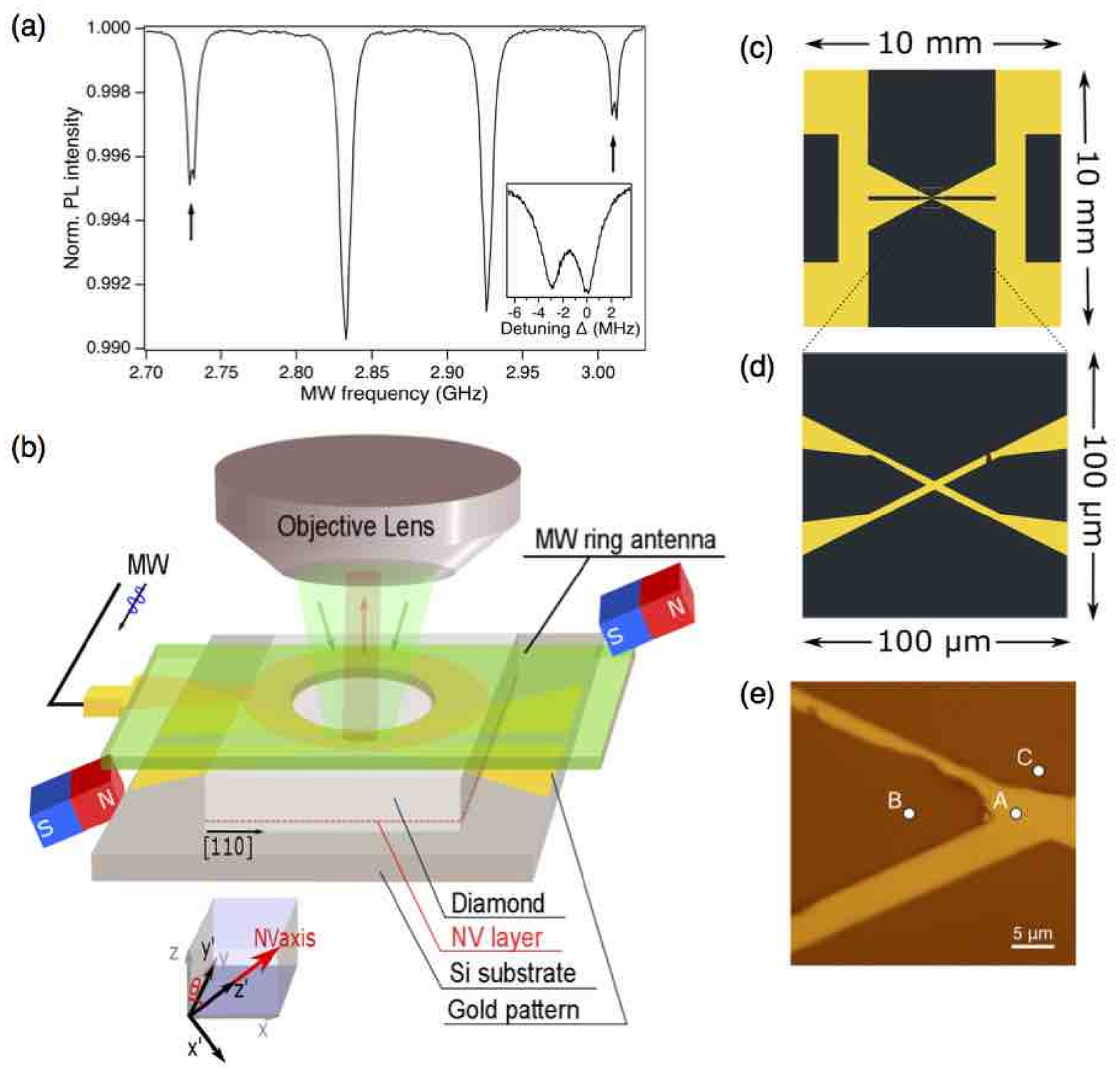}
\caption{\label{fig:expsetup} (a) Pulsed-ODMR spectrum of NV centers acquired at MW power 14.3 dBm and $\pi$-pulse duration of 1 $\mu$s with static magnetic field $B_{0}$ = 4.6 mT aligned along [111] direction. The two black arrows indicates the transitions $\ket{0}$ $\rightarrow$ $\ket{\pm1}$ correspondent to [111] direction used in this experiment to perform the MW magnetic field imaging. The two inner resonance peaks correspond to other three possible orientation for the N-V pair. The inset shows a zoom of the peak at 3.010 GHz which shows a double splitting associated with the hyperfine interaction of $^{15}$N nuclear spin at MW power of 9.3 dBm and MW pulse duration 1.5 $\mu$s. (b) Experimental setup used for the excitation of the micrometer-scale gold pattern and the imaging of the MW magnetic field distribution. The green laser at $\lambda$ = 520 nm is focused through an objective lens 100x on the NV layer. (c) Sketch in the xy-plane of gold pattern with thickness 110 nm coated on silicon substrate with size of 0.5 x 10 x 10 mm$^3$. (d) Zoomed area with size of 100 x 100 $\mu$m$^{2}$ showing the crossing point of micrometer-scale gold pattern. (e) Optical image of the area of measurement, correspond to the highest amplitude of the MW field. The points A, B, C indicate the position of the Rabi oscillations shown in Fig.\ref{fig:RabiBulk}a, b and c.}
\end{figure}

An NV center in diamond is a defect which constitutes of a vacancy in the diamond lattice and a near nitrogen atom which substitutes a carbon atom. The ground state of NV centers is a spin-triplet whose single state m$_{s}$ = 0 and doublet state m$_{s}$ = $\pm$1, named here $\ket{0}$ and $\ket{\pm 1}$, has a transition (zero-field splitting) at frequency 2.87 GHz which makes them ideal for imaging microwave fields. The degeneracy of the states $ \ket{\pm1}$ is lifted by a static external field $B_{0}$ which produces a Zeeman energy splitting of 2$\gamma$$B_{0}$, where $\gamma$ is the gyromagnetic ratio of the electron spin. The spin state can be optically initialized by off-resonance green laser pumping in the state $\ket{0}$. After manipulation with a MW field resonant with the transitions $\ket{0}$ $\rightarrow$ $\ket{\pm1}$, the spin state is measured by optically detected magnetic resonance (ODMR) through photoluminescence emitted in 600$-$800 nm range. The static magnetic field $B_{0}$ employed to remove the degeneracy of states $\ket{\pm 1}$ produces eight magnetic resonances, correspondent to the four possible orientations of NV-center and two spin transition $\ket{0}$ $\rightarrow$ $\ket{\pm1}$. In principle, measuring the projection of the microwave field resonance employing all the possible N-V directions ([111], [-111], [1-11], [-1-11]), it would be possible to fully reconstruct the MW magnetic field \cite{Maertz2010, Steinert2010}. Here, we align $B_{0}$ = 4.6 mT along the [111] direction, whose resonance transition will be used to map the MW field distribution on the micrometer-scale gold pattern. In this case, the projection of $B_{0}$ along the other possible NV directions is the same, leading to four magnetic resonances appear in the ODMR spectrum, two for the spin transition of [111] and two for the other directions, as shown in Fig.\ref{fig:expsetup}a. The hyperfine interaction between the electronic spin and the $^{15}$N nuclear spin ($I=1/2$) causes a further double energy splitting $A_{||}=3$ MHz, measured for the peak centered at 3.010 GHz in the ODMR spectrum (See e.g. Rabeau et. al. 2006 \cite{Rabeau2006}).
\section{}
The core of our experimental setup is a (100) CVD type IIa ultra-pure diamond substrate with the size of 2.0 x 2.0 x 0.5 mm$^3$, employed as platform for MW field imaging. After implantation of $^{15}$N$_{2}^{+}$ ions at 10keV with dose $2$ x $10^{13} $cm$^{-2}$\cite{Okai2012}, the diamond substrate is annealed at 800$^{\circ}$C and treated by acid. The ensemble of NV centers created after the annealing is located at a depth of $\sim$10 nm from the surface. Figure \ref{fig:expsetup}b shows a pictorial representation of our experimental setup. The diamond substrate is cut along [110] direction, aligned along the x-direction in the laboratory frame. The [111] direction is tilted respect to the z-axis at an angle $\theta$ = 54.74$^{\circ}$ in the xz-plane. The diamond substrate is sandwiched between a planar ring MW antenna and a micrometer-scale gold pattern fabricated on a silicon substrate. The MW field generated by the planar ring antenna excites the gold pattern which redistributes the microwave magnetic field, enhancing it in a localized area. Figure \ref{fig:expsetup}c and d, show a sketch of the micrometer-scale gold pattern and in Fig.\ref{fig:expsetup}e it's shown an image of the area of measurement taken by optical microscope. The planar ring MW antenna is a single-loop coil surrounding a circular hole with radius 0.5 mm and provides a spatially uniform magnetic field in an area of $0.785$ mm$^2$; low return losses S$_{11}$ in a range 2.7$-$3.2 GHz and input impedance of 50 $\Omega$ \cite{Sasaki16}. Microwave pulses were created by a signal generator (SMC100A, Rhodes Schwarz) and a microwave switch (ZFSWA2-63DR+, Mini-Circuits), and were amplified by a power amplifier (ZHL-16W-43+, Mini-Circuits). The gold structure is prepared by electron beam evaporation of 110 nm of gold on a silicon substrate with size of $10$ x $10$ x 0.5 mm$^3$. Note that the peculiarity of this experiment is that the micrometer-scale gold pattern acts as a receiving antenna and it’s not fed by any electrical cable. The NVs layer which will sense the MW magnetic field distribution faces the gold pattern, kept in contact with the surface of the diamond. The measurements are performed by means of a confocal microscope and a pulsed laser diode with $\lambda$ = 520 nm driven by a high speed driver at the peak power of 70 mW used to excite the NV centers. The photoluminescence arising from the diamond substrate is collected by an objective lens 100x NA 0.73 WD 4.7 mm with resolution of 0.5 $\mu$m at $\lambda$ = 700 nm; then the light is focused into a cooled scientific CMOS camera, 528 x 512 pixels. The static magnetic field to remove the degeneracy of the states $\ket{\pm 1}$ is provided by two Nd$_{2}$Fe$_{14}$B permanent magnets aligned along the [111] direction.
\section{}
The ability of the non-fed micrometer-scale gold pattern in reshaping the MW field generated by the antenna, provides a powerful tool for specific spin manipulations in diamond. In order to show the characteristics of this structure, we performed the imaging of the MW magnetic field distribution over the micrometer-scale gold pattern, by using a dense ensemble of NV centers in diamond.  According to the selection rules, the transition between the states $\ket{0}$ and $\ket{\pm1}$ is sensitive to circular polarization \cite{Alegre2007} and in a Rotating Wave Approximation (RWA) is allowed only for circularly polarized microwave fields. The MW field to be imaged, in resonance with the spin transitions $\ket{0}$ $\rightarrow$ $\ket{\pm 1}$, drives Rabi oscillations between the two levels with Rabi frequency $\Omega_{0} / 2\pi=\gamma B_{\pm}$, where $\gamma$ = 28 GHz/T is the electronic gyromagnetic ratio and $B_{+(-)}$ is amplitude of the left (right) handed circularly polarized amplitude of the microwave field $\sigma_{\pm}$. The Rabi frequency is directly proportional to amplitude of the magnetic field and the imaging of the MW field at a certain position can be performed by measuring the Rabi oscillations and calculating the relative frequency by Fast Fourier Transform (FFT). The components of the MW field in the perpendicular plane of the N-V axis, drives the Rabi oscillations of the electron spin and are thus sensed by the system.
\begin{figure*}[!ht]
\includegraphics[width=\linewidth]{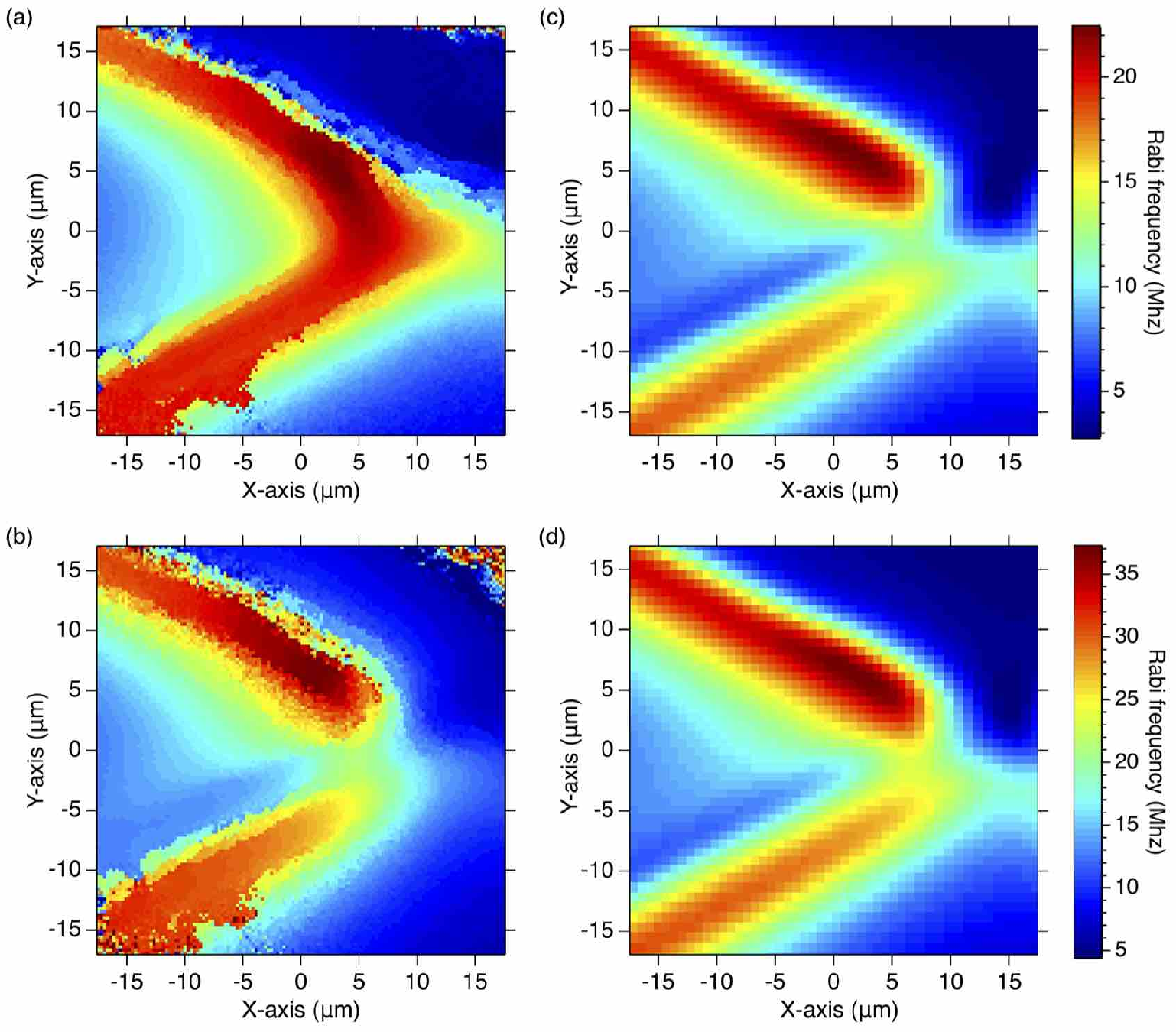}
\caption{\label{fig:expsim}Comparison between the experimental results of the Rabi frequency distribution on the micrometer-scale gold pattern at a MW power of 29.3 dBm and the relative simulation performed by FDTD analysis. (a) Imaging of the MW field with frequency 3.010 GHz, resonant with the transition $\ket{0}$ $\rightarrow$ $\ket{+1}$ and its simulation in (c). (b) Imaging of the MW field with frequency 2.730 GHz, resonant with the transition $\ket{0}$ $\rightarrow$ $\ket{-1}$ and its simulation (d). The MW distribution in the near-field calculated by the simulation is taken at a distance of 3.4 $\mu$m above the gold pattern. The gold pattern is excited by a double dipole antenna located at distance of 30 mm above the gold surface. The dipoles generate a linearly polarized magnetic field perpendicular to the pattern. After the simulation, the MW distribution is calculated by aligning the z-axis in the laboratory frame of reference along the [111] direction and taking the circularly polarized MW field in the perpendicular plane of as explained in the text.}
\end{figure*}
The external magnetic field $B_{0}$ should be strong enough to separate the resonance at [111] from the resonances correspondent to other directions. In fact, for a large MW powers, the MW field resonant with [111] could drive Rabi oscillations of close resonances, introducing multiple components at higher frequency. In the following experimental results, we choose 37.3 dBm as upper limit for the MW power since we started to observe these components in the FFT spectrum.
 
The Rabi oscillations are measured by first applying a green laser pulse with duration 1 $\mu$s which prepares the spin in the $\ket{0}$ state. After a waiting time of 1 $\mu$s necessary to complete the spin polarization, the electronic spin is driven with a MW pulse resonant with $\ket{0}$ $\rightarrow$ $\ket{\pm 1}$; the spin state is measured by applying a second laser pulse with the same duration, which reinitializes the spin state in $\ket{0}$. The CMOS camera acquires images in area with size of 34.9 x 33.9 $\mu$m$^{2}$ at increasing MW pulse durations to reconstruct temporally the Rabi oscillations. For a fixed MW pulse duration the previous sequence is repeated 24000 Cycles, correspondent to the integration time of the CMOS camera; at the end, the single measurement is repeated 200 times and averaged. The signal is normalized by acquiring the same sequence with MW pulse off. The Rabi oscillations at a specific position are calculated by averaging an area of 4 x 4 pixels in the images acquired by the CMOS camera. On one hand, the average improve the signal to noise ratio, mitigating movements of experimental setup during a long measurement. On the other hand, it limits the resolution of the measurement to an area of size 264 x 264 nm$^{2}$ but smaller than the minimum width of the gold pattern ($\sim$1 $\mu$m) and enough to reconstruct the main features of the MW field distribution. The frequency of the Rabi oscillations of 4 x 4 pixels is calculated performing the FFT and identifying the maximum intensity peak in the spectrum. 

Note that the double splitting related with the hyperfine interaction with the $^{15}$N nuclear spin, as shown in the inset of Fig.\ref{fig:expsetup}a, suggest that we have to select one the two resonances. Considering one of the two peaks, an off-resonance MW field would drive Rabi oscillations with an effective Rabi frequency $\Omega_{eff}$=$\sqrt{\Omega_{0}^{2}+\Delta^{2}}$, where $\Omega_{0}$ is the Rabi frequency on-resonance and  $\Delta$ the detuning off-resonance. This means that a MW field resonant with one of two transitions would cause a double frequency to appear in the FFT spectrum of Rabi oscillations for frequencies close to $A_{||}=3$ MHz. A simple and effective solution is to choose the MW frequency at the center of the nuclear spin splitting with $\Delta$ $ \simeq$ 1.5 MHz thus taking the same contribution in frequency detuning from the two resonances and avoiding beats in the Rabi oscillations. In the case of $^{14}$N isotopes with three resonance peaks, this is not possible and a beat would appear in the Rabi oscillations (see e.g. Wang \latin{et. al.} \cite{Wang2018} ). However, in this way the minimum Rabi frequency sensed in our measurements is limited by  $\Delta$ and only for $\Omega_{0}^{2}\gg\Delta^{2}$ we can assume $\Omega_{eff}\simeq\Omega_{0} = 2\pi\gamma$B$_{\pm}$ an thus the linear relation between the MW magnetic field and the effective Rabi frequency.
\section{}
We performed the imaging of the MW magnetic field distribution at frequency of 2.730 GHz for $\ket{0}$ $\rightarrow$ $\ket{-1}$ and at frequency of 3.010 GHz for $\ket{0}$ $\rightarrow$ $ \ket{+1}$, for different powers fed to our planar ring MW antenna. The imaged area of the MW field that we chose to report here is the central part of the crossed pattern as shown in Fig.\ref{fig:expsetup}e, corresponding to the maximum value of the enhanced MW field. Figure \ref{fig:expsim} shows the imaging of the Rabi frequency distribution measured at a MW power of 29.3 dBm for the $\ket{0}$ $\rightarrow$ $\ket{\pm1}$ transition; the experimental result is compared with a numerical simulation performed by finite-difference time-domain (FDTD) analysis, using the open source program “OpenFDTD” \cite{OpenFDTD}. In the simulation, the source of MW field is a double dipole antenna phase-shifted 180$^{\circ}$ which ensures a linearly polarized field along the direction perpendicular to the gold pattern (z-axis, in the lab frame). For a more accurate reconstruction of the MW field distribution, we reproduced the shape of the gold pattern according to the optical images, including main asymmetries and imperfections. The MW magnetic field as measured by NV centers, is calculated by first transforming of the frame of reference of the laboratory  $\left\{ x,y,z \right\}$ in a new set of coordinates $\left\{ x',y',z' \right\}$ as shown in Fig.\ref{fig:expsetup}, where z$'$-axis is aligned along the NV-axis at [111] direction. The circularly polarized amplitude of the MW field $\sigma_{\pm}$ sensed by NV centers can be calculated with the following formula:

\begin{equation}\label{eq:pol}
B_{\pm}=\abs{B\hat{x'} \mp B\hat{y'}}
\end{equation}

where $+ \left( -\right)$ is the left (right) handed component of the circularly polarized light and $\hat{x'}$ and $\hat{y'}$ are the unit vectors in the new N-V frame. We kept in contact the surfaces of diamond and gold in our experimental setup but since a small gap $\left(\sim\mu m \right)$ might exist, among our simulation results we chose the most representative frequency distribution at a distance of 3.4 $\mu$m from the micrometer-scale gold pattern. The measurements in Fig.\ref{fig:expsim}a, b show a maximum Rabi frequency of 22.54 MHz (37.18 MHz) above the gold pattern for the transition $\ket{0}$ $\rightarrow$ $\ket{+(-)1}$, decreasing gradually away from it. The asymmetries in the Rabi frequency distribution are due to imperfections in the structure of the gold pattern. Note that these imperfections produce different MW distributions for the two spin transitions. This suggests that the asymmetries cause the MW field emitted by the gold pattern to be partially circularly polarized. The simulated distributions in Fig.\ref{fig:expsim}c, d can reproduce the main features of the measurements except the different patterns according to the spin transition. We explained this disagreement with the fact that some smaller areas of the gold pattern and imperfections are not well reproduced due to the limited mesh size of the FDTD program and we didn't further investigate. We stress again that fact that the gold pattern is excited by the planar ring MW antenna at a distance of $\sim$1 mm, without a direct feed. Moreover, for the particular shape of the pattern employed in this experiment, the receiving gold pattern produced a localized MW field on the micrometer-scale, compared to the bulk MW field generated by the only planar ring MW antenna on the millimeter-scale. 
\\
\begin{figure*}[!ht]
\includegraphics[width=\linewidth]{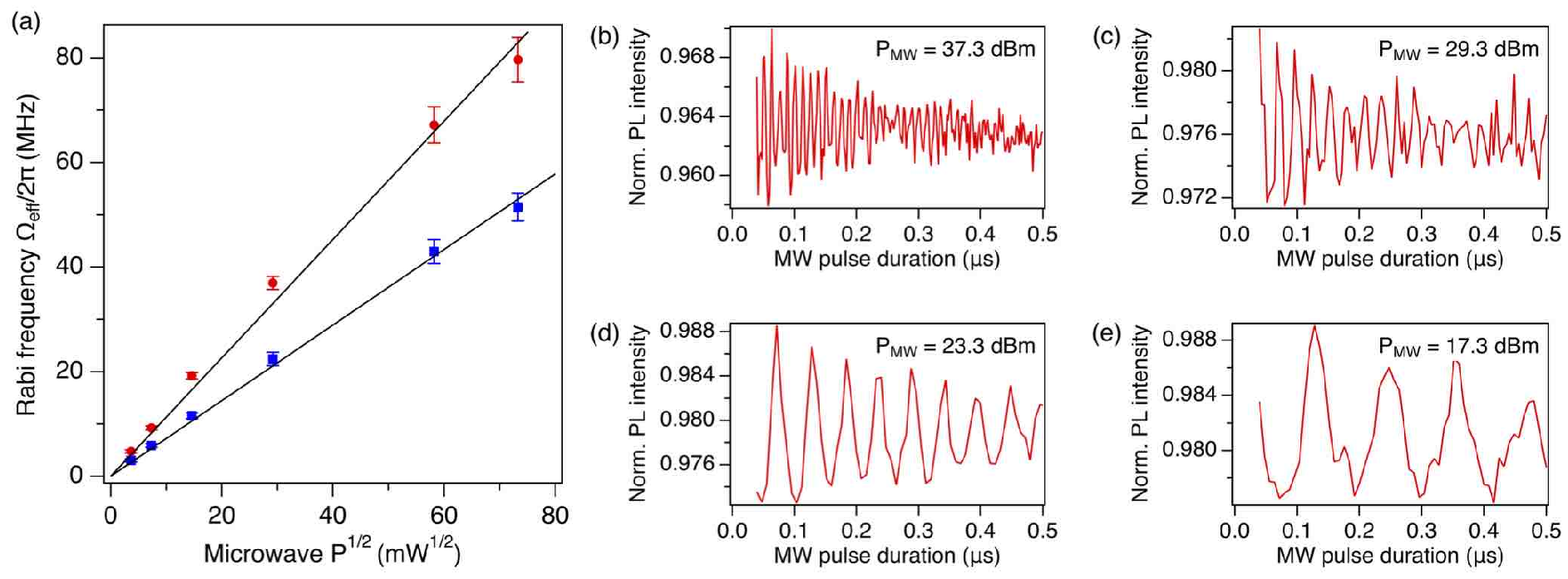}
\caption{\label{fig:RabiLin} (a) Dependence of the effective Rabi frequency $\Omega_{eff}$ on the square root of the MW power fed to the planar ring MW antenna for the transition $\ket{0}$ $\rightarrow$ $\ket{-1}$ (red circles) and $\ket{0}$ $\rightarrow$ $\ket{+1}$ (blue squares). Considering the images of the Rabi frequency distribution with size 132 x 128 pixels, the values are calculated by averaging the first 8 pixels near the maximum value for an area of size $\sim$0.63 $\mu$m$^2$. The error bars are calculated from the standard deviation of the previous average and summed to the half width half maximum (HWHM) value of the peak in the FFT spectrum. The black line is a linear fit which show the linear dependence of the effective Rabi frequency on the square root of the MW power. The right panel shows the Rabi oscillations extracted from (a), measured in the point with maximum frequency above the gold pattern for the transition $\ket{0}$ $\rightarrow$ $\ket{-1}$ at MW power: (b) 37.3 dBm, (c) 29.3 dBm, (d) 23.3 dBm and (e) 17.3 dBm. }
\end{figure*}
At the same position, we measured the maximum Rabi frequency for different MW powers (11.3 $-$ 37.3 dBm). Figure \ref{fig:RabiLin}a depicts the linear dependence of the Rabi frequency on the square root of the MW power $\sqrt{P_{MW}}$ of our antenna. As explained previously, even if we measure an effective Rabi frequency, the linear relation is preserved in the limit of fast enough Rabi oscillations. The different slopes for the transition $\ket{0}$ $\rightarrow$ $\ket{\pm1}$ is due to a different gain of the MW antenna at a frequency of the transition. In Fig.\ref{fig:RabiLin}b-d we compare the Rabi oscillations measured at different MW power for the transition $\ket{0}$ $\rightarrow$ $\ket{-1}$. The incresing decay for higher powers is mainly due to the random fluctuations in the power of the MW source \cite{Fedder2011} and this problem could be overcome by applying a decoupling sequence such as the Concatenated Continuos Driving (CCD) scheme \cite{Cai2012}. 

\begin{figure*}[!ht]
\includegraphics[width=\linewidth]{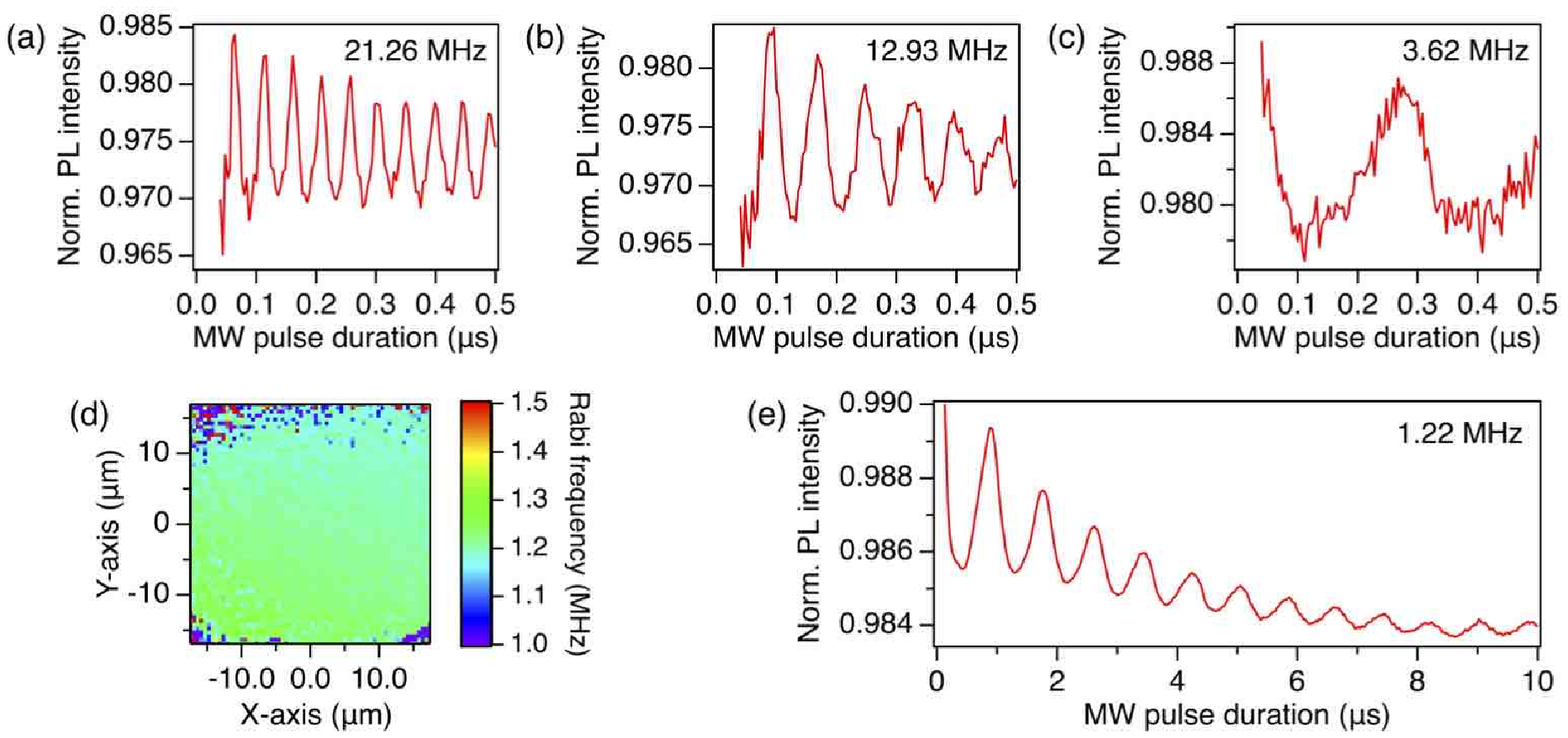}
\caption{\label{fig:RabiBulk} (a-c) Rabi oscillations averaging 4 x 4 pixels in area with size 264 x 264 nm$^2$ acquired at MW power of 29.3 dBm for the transition $\ket{0}$ $\rightarrow$ $\ket{+1}$, at different position over the micrometer-scale gold pattern, as indicated by the white circles in Fig.\ref{fig:expsetup}e. (d) Imaging of the bulk Rabi frequency distribution far from the micrometer-scale gold pattern, on-resonance with one of the peaks originated by the hyperfine interaction with $^{15}$N nuclear spin. The position of the FFT peak is determined after a zero-filling of 10 times the measured Rabi oscillations to resolve a minimum shift of 10 kHz. (e) Bulk Rabi oscillations with frequency $\Omega_{0}$/$2\pi$ = 1.22 MHz calculated by averaging the area in (d).}
\end{figure*}

In order to have a quantitative information about the enhancement of the MW magnetic field on the micrometer-scale gold pattern, we imaged the MW field distribution at 29.3dBm for the transition $\ket{0}$ $\rightarrow$ $\ket{+1}$ far from the micro-scale gold pattern, which correspond to Rabi oscillations in presence of the only MW planar ring antenna. Since the frequency of the oscillations is comparable with the nuclear spin splitting, we performed the measurement of Rabi oscillations for a frequency detuning $\Delta$ = 0 (see the inset of Fig.$\ref{fig:expsetup}$), resonant with one nuclear spin transition to clearly determine the Rabi frequency. As it's shown in Fig.\ref{fig:RabiBulk}d, the distribution of the Rabi frequency is homogeneous with frequency $\Omega_{0}/2\pi\simeq$ 1.22 $\pm$ 0.04 MHz  in the area of imaging. Comparing this result with the maximum field measured in Fig.\ref{fig:expsim}a, we can estimate a MW magnetic field enhancement of about 19 times larger for the micro-scale gold pattern, localized in a micrometer-scale area.
\section{}
We have demonstrated the enhancement of a MW magnetic field generated by a planar ring MW antenna by means of a resonant excitation of a micrometer-scale gold pattern, without a direct feed of electrical current. We have measured the distribution of the MW magnetic field at a frequency of 2.730 GHz and 3.010 GHz by driving the Rabi oscillations of the electronic spin in NV centers in diamond, showing the NV centers as a valuable method for direct imaging of MW fields. The Rabi frequency directly proportional to MW magnetic field amplitude let us evaluate the MW distribution by calculating the Rabi frequency distribution over the micrometer-scale gold pattern. According to our experimental results, the micrometer-scale gold pattern creates a maximum Rabi frequency enhanced 19 times compared to the Rabi frequency in presence of the single MW planar ring antenna. The enhanced MW field is localized in area on the scale of the minimum width of the gold pattern ($\sim$1 $\mu$m) compared to MW field generated by the planar ring MW antenna, distributed in an area of size 0.785 mm$^{2}$. We believe that our micrometer-scale gold pattern would be a powerful tool for the coherent manipulation of few single spins in nanometer-scale for quantum information applications.
\begin{acknowledgement}
This work was partly supported by a Grant-in-Aid for Scientific Research (Nos. JP15H05853,  JP18H04283, and JP18K18726) from Japan Society for the Promotion of Science.
\end{acknowledgement}
\clearpage

\bibliography{achemso-demo}

\end{document}